\documentclass[%
 aps,
 prl,
amsmath,amssymb,
reprint,
]{revtex4-1}
\usepackage{graphicx}% Include figure files
\usepackage{dcolumn}% Align table columns on decimal point
\usepackage{bm}% bold math
\usepackage[utf8]{inputenc}
\usepackage{mathptmx}
\usepackage{multirow}
\usepackage{braket}

\usepackage[english]{babel}
\usepackage{xcolor}% Include colors for document elements
\colorlet{RED}{red}
\colorlet{BLUE}{blue}
\usepackage{dcolumn}% Align table columns on decimal pointhttps://v1.overleaf.com/17481932tmwjwzytcbcq#
\usepackage{bm}% bold math
\usepackage[version=4]{mhchem} % HvD: for general chemical formula like \ce{(TiO2)n}
\usepackage{acronym}
\usepackage{adjustbox}
\usepackage{float}
\usepackage{tikz}
\usepackage{microtype} % fixes many of the overflow hbox errors
\usepackage{algorithm}
\usepackage{xcolor}
\usepackage{algpseudocode}

\usetikzlibrary{calc,shapes.geometric,decorations.pathmorphing,patterns}

\definecolor{background-color}{gray}{0.98}
\usepackage[margin=2.3cm,bmargin=1cm,footnotesep=1cm]{geometry}

\begin{document}

\title{
%Searching a compact unitary basis for the quantum simulation of non-unitary 
Quantum flow algorithms for simulating  many-body systems on quantum computers
}

\author{Karol Kowalski}
\email{karol.kowalski@pnnl.gov}
\affiliation{%
  Physical Sciences Division, 
  Pacific Northwest National Laboratory, Richland, Washington, 99354, USA
}

\author{Nicholas P. Bauman}
\affiliation{%
  Physical Sciences Division, 
  Pacific Northwest National Laboratory, Richland, Washington, 99354, USA
}

\begin{abstract}
% We report  quantum simulations of strongly correlated systems  via the quantum flow (QFlow) approach, which enables sampling large sub-spaces of Hilbert space by  coupled  eigenvalue problems defined in reduced dimensionality active spaces. 
% We design QFlow algorithms that  significantly reduce the circuit complexity, and offer a way towards the scalable and constant-circuit-depth realization of quantum computing. Our simulations for model systems indicate that QFlow can optimize the number of wave function parameters 
% that is more than an order of magnitude larger than the number of parameters for a single active space without an increase in the required  qubits. 

% 1.) iterative process 1 sentence (amplitudes to build effective Hamiltonians
% are taken from previous iterations/cycles
% 2.)number of parameters: discussion
% 3.) simplest model for downfolded Hamiltonian A(6)-A(7) one- and two-body terms
% 4.) in Conclusions: QFlows for quantum dynamics
% Design theory to take advantage of currently exustung quantum computers

We conducted quantum simulations of strongly correlated systems using the quantum flow (QFlow) approach, which enables sampling large sub-spaces of the Hilbert space through coupled variational problems in reduced dimensionality active spaces. Our QFlow algorithms significantly reduce circuit complexity and pave the way for scalable and constant-circuit-depth quantum computing. Our simulations show that QFlow can optimize the collective number of wave function parameters without increasing the required qubits using active spaces having an order of magnitude fewer number of parameters.

\end{abstract}

\maketitle

\paragraph{Introduction.---} 
The development of quantum computing has grabbed the attention of the many-body chemistry and physics communities with the promise to provide exponential speed-ups over traditional computing for problems such as solving the electronic Schrödinger equation for ground and excited states or the time-dependent equation for studying dynamics. For the electronic problem, the two salient quantum algorithms for determining energetics with the electronic Hamiltonian are quantum phase estimation (QPE)\cite{nielsen2002quantum,kitaev1995quantum,Kitaev_1997,Abrams1999QPE,childs2010relationship,reiher2017elucidating} and variational quantum eigensolver (VQE) \cite{
peruzzo2014variational,mcclean2016theory,romero2018strategies,Kandala2017,kandala2019,izmaylov2019unitary,lang2020unitary,grimsley2019adaptive,grimsley2019trotterized,mcardle2020quantum,Love2021,tilly2022variational}. Both algorithms are seized by complexities that prevent routine calculations of meaningful problems that plague traditional computing. These complexities result from the inherently large dimensionality needed to provide accurate and reliable results. For QPE, this manifests in circuit depths far beyond what is achievable in the noisy intermediate-scale quantum (NISQ) device era of quantum computing. For VQE, a measure of complexity is the number of parameters from a given ansatz currently optimized using traditional computing algorithms. 
The progress in enabling quantum computing technologies is contingent not only on the advances in the design of quantum materials but also on the ability to adapt to new methodological advances in the theory of correlated many-body systems.

The reduction of dimensionality and compression of quantum Hamiltonians has become a crucial area of focus in the realm of quantum computing. In light of this, it is of utmost importance to develop methodologies that aim to compress the correlation effects in smaller spaces that can be handled by current quantum computing resources
\cite{Tyler2020,csahinouglu2021hamiltonian,EffectiveH2022}.
As such, the authors have devised a coupled cluster (CC)-based downfolding formalism that enables the incorporation of dynamical correlation effects from large Hilbert spaces into manageable effective Hamiltonians for a smaller sub-space of the original problem.\cite{bauman2019downfolding}
This letter describes and provides numerical evidence for a new dimensionality-reducing technique called the quantum flow (QFlow) approach. 
%The QFlow algorithm is a hybrid computational workflow that integrates the most appealing quantum and classical computing features by 
%partitioning large sub-space of the Hilbert space into coupled eigenvalue problems in reduced dimensionality active spaces.\cite{kowalski2021dimensionality} 
The QFlow algorithm integrates the reduced-dimensionality active space variational problems to approximate the ground-state energy of the Hamiltonian operator within a larger sub-space of Hilbert space
\cite{kowalski2021dimensionality}.

Within the framework of QFlow formalism, the highest demand for qubits is linked to the number of qubits necessary for the representation of the quantum problem that corresponds to the largest  active space incorporated in the flow.
%dimensionality of a given problem is reduced to the dimensionality of the largest active space. 
Using modest-size active spaces, we demonstrate that QFlow can efficiently recover the corresponding energetics of the full problem.  It is a flexible workflow that we expected to play a pivotal role in performing quantum simulations 
on quantum computers during the transition from NISQ devices to fully-fledged error-corrected quantum computing.

\paragraph{CC  Theory and Quantum Flows.---} The CC theory 
\cite{coester58_421,coester60_477,cizek66_4256,paldus1972correlation,arponen83_311,bishop1987coupled,paldus07,bartlett_rmp} has evolved into a one of the most prominent formalisms to describe correlated systems. In the single-reference variant (SR-CC), the ground-state wave function $|\Psi\rangle$ is defined by the exponential Ansatz
% revision
%{\color{blue}
\begin{eqnarray}
&& |\Psi\rangle = e^T|\Phi\rangle \;,
\label{cca} \\
T &=& \sum_{k=1}^{N_A} \frac{1}{(k!)^2} 
\sum_{\substack{i_1,\ldots,i_k \\ a_1,\ldots,a_k}}
t^{a_1 \ldots a_k}_{i_1 \ldots i_k} 
a_{a_1}^{\dagger}\ldots a_{a_k}^{\dagger} a_{i_k} \ldots a_{i_1} \;,
\label{ccaa}
\end{eqnarray}
%} % end of revision
where $T$ and $|\Phi\rangle$ represent the cluster operator and reference function.
% revision
%{\color{blue}
The $T$ operator is defined by the maximum excitation level ($N_A$), cluster amplitudes $t^{a_1 \ldots a_k}_{i_1 \ldots i_k}$, and creation/annihilation operators $a_p^{\dagger}$/$a_q$ where $p,q$ stand for the general spin-orbital indices. The indices $i_j$ ($a_j$) stand for occupied (unoccupied) spin-orbitals in the reference function $|\Phi\rangle$.
%} % end of revision
Standard CC equations are given by the equations
\begin{eqnarray}
Qe^{-T}He^T |\Phi\rangle &=& 0 \;, \label{cceq} \\
\langle\Phi|e^{-T}He^T|\Phi\rangle &=& E\;, \label{ccene}
\end{eqnarray}
where $Q$ is a projection operator onto excited Slater determinants generated by acting with $T$ on $|\Phi\rangle$
(the projection onto the reference function is denoted as $P$). Recently, it has been demonstrated that CC energies can be calculated by diagonalizing effective Hamiltonians in  a class of complete active spaces (CASs) that are specific to the approximation of the $T$ operator\cite{kowalski2018properties,kowalski2021dimensionality,kowalski2023sub}. If, in the particle-hole formalism,  CAS is generated by the excitation sub-algebra ($\mathfrak{h}$), and the cluster operator $T$ can be partitioned into  internal 
($T_{\rm int}(\mathfrak{h})$; producing excitation within CAS) and external ($T_{\rm ext}(\mathfrak{h})$; producing excitation outside of  CAS) parts and $e^{T_{\rm int}(\mathfrak{h})}|\Phi\rangle$ represents an exact-type expansion in the CAS, then  the CC energy can be obtained as:
\begin{equation}
H^{\rm eff}(\mathfrak{h}) e^{T_{\rm int}(\mathfrak{h})}|\Phi\rangle =  E e^{T_{\rm int}(\mathfrak{h})}|\Phi\rangle  \;, \label{sescc1}
\end{equation}
\begin{equation}
H^{\rm eff}(\mathfrak{h})  = (P+Q_{\rm int}(\mathfrak{h}))
e^{-T_{\rm ext}(\mathfrak{h})}He^{T_{\rm ext}(\mathfrak{h})}
(P+Q_{\rm int}(\mathfrak{h}))\;,
\label{sescc2}
\end{equation}
where $Q_{\rm int}(\mathfrak{h})$ is a projection onto excited (with respect to $|\Phi\rangle$)  configurations in the CAS. The above property of the CC formalism (referred to as the CC downfolding)  is valid for {\em any type of sub-algebra $\mathfrak{h}$} (henceforth referred to as the sub-system embedding sub-algebras (SES)) described above. 
% revision
%{\color{blue}
The partitioning of the cluster operator into internal and external parts has been originally introduced in the context of the state-selective CC formalism in Refs.\cite{pnl93,piecuch1994state,adamowicz1998state,piecuch_molphys}. Although the SES Theorem (Eq.(\ref{sescc1})) and Equivalence Theorem 
({\em vide infra}) are based on the decomposition of cluster operator into internal/external parts, the possibility of calculating CC energies in an alternative way and integrating various active-space problems were proposed only recently \cite{kowalski2018properties,kowalski2021dimensionality}. 
%}
% end of revision
The invariance of the CC energy with respect to the choice of SES led to the concept of quantum flow and Equivalence Theorem \cite{kowalski2021dimensionality,bauman2022coupled}, which states that when several SES problems 
represented by (\ref{sescc2}) are coupled into the flow, i.e.,
\begin{equation}
H^{\rm eff}(\mathfrak{h}_i) e^{T_{\rm int}(\mathfrak{h}_i)}|\Phi\rangle =  E e^{T_{\rm int}(\mathfrak{h}_i)}|\Phi\rangle  \, (i=1,\ldots,M) \label{flow1}
\end{equation}
($M$ stands for the number of CASs included in the flow), the corresponding solution is equivalent to the standard representation of the CC theory given  by Eqs.~(\ref{cceq}) and (\ref{ccene}) with the $T$ operator defined as a 
combination of all 
non-repetitive
excitations included in
$T_{\rm int}(\mathfrak{h}_i)\; (i = 1, . . . , M)$ operators, 
% revision
%{\color{blue}
symbolically denoted as,
%}
%
\begin{equation}
  T= %\bigcup_{i=1}^{M} 
  {\widetilde{\sum}}_{i=1}^M 
  T_{\rm int}(\mathfrak{h}_i) \;.
  \label{flow2}
\end{equation}
An important consequence of the Equivalence Theorem is the fact that for some choices of cluster operator, Eq.~(\ref{flow2}), high-dimensionality problem,
Eqs.~(\ref{cceq}) and (\ref{ccene}) can be replaced by a flow composed of 
reduced-dimensionality non-Hermitian  eigenvalue problems. For each sub-algebra $\mathfrak{h}_i$ in the eigenvalue problem of Eq.~(\ref{flow1}), the effective Hamiltonian $H^{\rm eff}(\mathfrak{h}_i)$ follows Eq.~(\ref{sescc2}), where the external cluster operators $T_{\rm ext}(\mathfrak{h}_i)$ are the collection of operators excluding $T_{\rm int}(\mathfrak{h}_i)$,
\begin{equation}
  T_{\rm ext}(\mathfrak{h}_i) = T - T_{\rm int}(\mathfrak{h}_i) \;.
  \label{extTflow}
\end{equation}

% revision
%{\color{blue}
To extend the SR-CC flows to the Hermitian case, in contrast to previous analysis \cite{kowalski2021dimensionality},  we will employ variational principle using functional 
\begin{equation}
E = \langle\Phi| e^{-\sigma} H e^{\sigma} |\Phi\rangle
\label{var1}
\end{equation}
where a general-type  anti-Hermitian cluster operator $\sigma$ 
% revision
%{\color{blue}
($\sigma = \sum_{k=1}^{N_A} \frac{1}{(k!)^2} 
\sum_{\substack{p_1,\ldots,p_k \\ q_1,\ldots,q_k}}
\sigma^{q_1 \ldots q_k}_{p_1 \ldots p_k} 
a_{q_1}^{\dagger}\ldots a_{q_k}^{\dagger} a_{p_k}\ldots a_{p_1}$, $\sigma^{\dagger}=-\sigma$)
%}
% end of revision
includes all excitations needed to generate space that is 
too large to be handled by available quantum computers. 
To tackle the problem using limited quantum resources, let us assume that 
the $\sigma$ operator can be approximated by amplitudes included in anti-Hermitian operators $\sigma_{\rm int}(\mathfrak{h}_i) (i=1,\ldots, M$), producing excitations within corresponding active spaces ($AS(i)$) generated by sub-algebras $\mathfrak{h}_i \;$ 
\begin{equation}
  \sigma \simeq %\bigcup_{i=1}^{M} 
  {\widetilde{\sum}}_{i=1}^M 
  \sigma_{\rm int}(\mathfrak{h}_i) \;.
  \label{var2}
\end{equation}
%We will also order active spaces according to their importance, starting from $AS(1)$ encapsulating most of the correlation effects and ending at $AS(M)$, which contains the least amount of correlation effects. 
In the next step, we will look at the problem (\ref{var1}) from the point of view of $i$-th active space and decompose $\sigma$ operator as 
\begin{equation}
 \sigma \simeq \sigma_{\rm int}(\mathfrak{h}_i)+ \sigma_{\rm ext}(\mathfrak{h}_i) \;,
 \label{var3}
\end{equation}
and 
\begin{equation}
E= \langle\Phi| e^{-\sigma_{\rm int}(\mathfrak{h}_i) - \sigma_{\rm ext}(\mathfrak{h}_i)} H 
e^{\sigma_{\rm int}(\mathfrak{h}_i) + \sigma_{\rm ext}(\mathfrak{h}_i)}|\Phi\rangle \;.
\label{var4}
\end{equation}
Next,  we will utilize the order-$N$ active-space-specific Trotter formula (in analogy to Ref. \cite{kowalski2021dimensionality}) to expand exponents, which introduce active-space-specific $E(\mathfrak{h}_i)$ approximation to energy $E$: 
\begin{equation}
E(\mathfrak{h}_i) = \langle\Psi_{\rm int}(\mathfrak{h}_i,N)|H^{\rm eff}(\mathfrak{h}_i,N)|\Psi_{\rm int}(\mathfrak{h}_i,N)\rangle \;,\label{var5}
\end{equation}
where 
\begin{equation}
H^{\rm eff}(\mathfrak{h}_i,N)=(P+Q_{\rm int}(\mathfrak{h}_i)) \lbrack G^{(N)}_i\rbrack ^{-1} H G^{(N)}_i  (P+Q_{\rm int}(\mathfrak{h}_i))  
\label{var6} 
\end{equation}
\begin{equation}
    G^{(N)}_i=( e^{\sigma_{\rm ext}(\mathfrak{h}_i)/N}
     e^{\sigma_{\rm int}(\mathfrak{h}_i)/N})^{N-1} e^{\sigma_{\rm ext}(\mathfrak{h}_i)/N}
     \label{var7}
\end{equation}
and 
\begin{equation}
|\Psi_{\rm int}(\mathfrak{h}_i,N)\rangle = e^{\sigma_{\rm int}(\mathfrak{h}_i)/N} |\Phi\rangle \;.
\label{var8}
\end{equation}
The coupled variational problems (\ref{var5}) for $i=1,\ldots,M$ define the QFlow algorithm. As in the non-Hermitian case, the total pool of amplitudes optimized in the QFlow corresponds to all non-repetitive  amplitudes for active spaces included in the flow. As a consequence of the non-commutativity of operators defining 
$\sigma$-operators and the need to use Trotter approximations, the energy values  in (\ref{var5}) may be, in general  different. For this purpose, we introduce physically motivated ordering of the active spaces (the first (last), or primary  active space contains the most (the least) important part of correlation effects) based, for example, on the orbital energy criteria and use the energy ($E(\mathfrak{h}_1)$) to probe the energy in the QFlow procedure. 
The advantage of the quantum version of the QFlow algorithm stems from the fact that the qubits requirement is associated with the qubits requirement of the  largest active space in the flow.  In this Letter, we will mainly focus our attention on the simplest $N=1$ case where sets of amplitudes defining effective Hamiltonian and 
$|\Psi_{\rm int}(\mathfrak{h}_i,N=1)\rangle\equiv |\Psi_{\rm int}(\mathfrak{h}_i)\rangle$ are disjoint.

\begin{figure}
    \centering
    \includegraphics[width=0.8\linewidth]{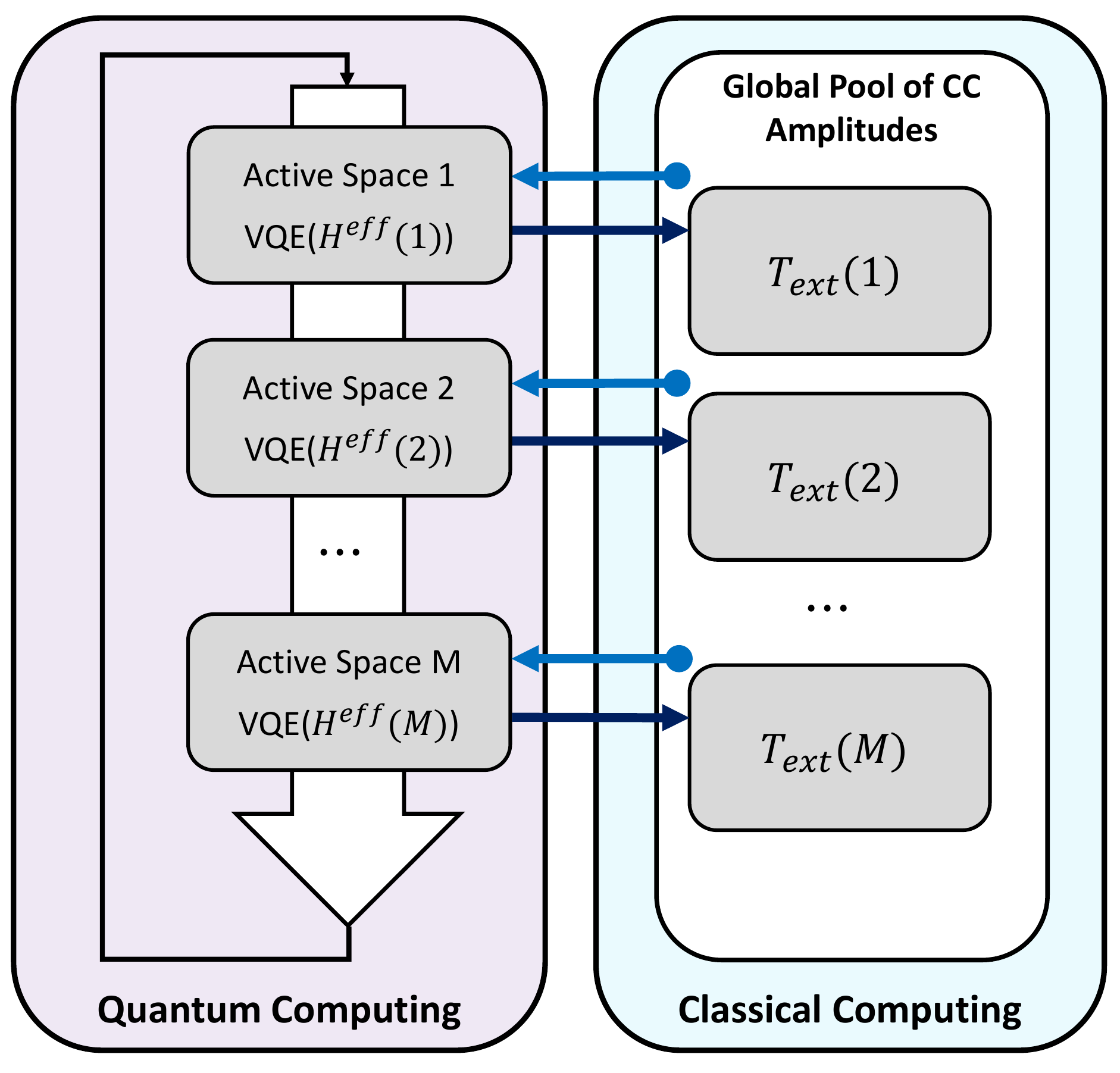}
    \caption{Schematic representation of the QFlow algorithms. All cluster amplitudes (Global Pool of CC amplitudes; (GPA)) are residing on classical computers. The effective Hamiltonians are formed on classical computers using GPA and encoded on quantum computers (light blue arrows). Quantum computers use these Hamiltonians to optimize internal excitations for a given active space and are used to update GPAs (dark blue lines).
}
    \label{fig:qflow}
%\end{figure*}
\end{figure}

\paragraph{Numerical Implementation.---}
The QFlow algorithm has never been validated numerically. To fill this gap, we developed QFlow  implementation for 
%revision
%{\color{blue} 
the first-order Trotterization approximation ($N=1$)  
%}
based on the stringMB code  - an occupation number representation-based emulator of quantum computing \cite{kowalski2023sub}. Our QFlow implementation, which emulates a VQE solver for each CAS involved in the flow, is schematically shown in Fig.~\ref{fig:qflow} and uses conventional computers to store a global pool of amplitudes and prepare effective Hamiltonian for each cycle. When representing the flow   in the form of  coupled eigenvalue problems, one has the flexibility in defining the active spaces, which can include those that overlap with each other and share common parameters. For the $\mathfrak{h}_i$ computational block we partition the set of variational  parameters 
${\bm \theta}(\mathfrak{h}_i)$ into subset ${\bm \theta}^{\rm CP}(\mathfrak{h}_i)$ that refers to  common pool of  
amplitudes determined in preceding steps (say, for $\mathfrak{h}_j \; (j=1,\ldots,i-1)$) and subset  ${\bm \theta}^{\rm X}(\mathfrak{h}_i)$ that is uniquely determined in the $\mathfrak{h}_i$ minimization step for $E(\mathfrak{h}_i)$, i.e, 
 \begin{widetext}
\begin{equation}
\min_{{\bm \theta}^{\rm X}(\mathfrak{h}_i)}   
\langle\Psi_{\rm int}({\bm \theta}^{\rm X}(\mathfrak{h}_i),{\bm \theta}^{\rm CP}(\mathfrak{h}_i))
|
H^{\rm eff}(\mathfrak{h}_i)
|
\Psi_{\rm int}({\bm \theta}^{\rm X}(\mathfrak{h}_i),{\bm \theta}^{\rm CP}(\mathfrak{h}_i))\rangle \;,\;(i=1,\ldots,M)\;,
\label{ducc5}
\end{equation}
\end{widetext}
where 
%{\color{blue}
$H^{\rm eff}(\mathfrak{h}_i)$ $\equiv$
$H^{\rm eff}(\mathfrak{h}_i,N=1)$
%} 
and $|\Psi_{\rm int}({\bm \theta}^{\rm X}(\mathfrak{h}_i),{\bm \theta}^{\rm CP}(\mathfrak{h}_i))\rangle$ 
(chosen in the form of unitary CC (UCC) Ansatz \cite{unitary1,unitary2,evangelista2011a}) approximates 
$|\Psi_{\rm int}(\mathfrak{h}_i)\rangle=e^{\sigma_{\rm int}(\mathfrak{h}_i)} |\Phi\rangle$ in Eq.~(\ref{var8}). When combined with  a simple form of the  gradients estimates on the  quantum computers \cite{grimsley2019adaptive}
\begin{equation}
\frac{\partial E(\mathfrak{h}_i)}{\partial \theta^{\rm X}(\mathfrak{h}_i)_k} \simeq
\langle
\Psi_{\rm int}(\mathfrak{h}_i)|
[H^{\rm eff}(\mathfrak{h}_i),\tau^X_k(i)]
|\Psi_{\rm int}(\mathfrak{h}_i)\rangle \;,
\label{ducc6}
\end{equation}
where $\tau^X_k(i)$ is a corresponding combination of the strings of a creation/annihilation operators associated with the  $\theta^{\rm X}(\mathfrak{h}_i)_k$ amplitude in the $\sigma_{\rm int}(\mathfrak{h}_i)$ operator. Instead of performing full optimization for each active space included in the QFlow, we perform only one optimization step based on the gradient
(\ref{ducc6}). We also employ UCC-type representation for each 
$\sigma_{\rm ext}(\mathfrak{h}_i)$ needed to construct $H^{\rm eff}(\mathfrak{h}_i)$ operator in Eq.~(\ref{var6}).
\paragraph{Results.---}
As a test system to demonstrate the performance of the QFlow techniques, we chose the H$_n$ linear chains of the hydrogen atoms: H6 and H8 models in small STO-3G basis set \cite{hehre1969self}, where one can vary the complexity of the ground-state wave function  by changing the H-H distances ($R_{\rm H-H}$) between adjacent atoms. For example, while for $R_{\rm H-H}$= 2.0 a.u., one deals with the weakly correlated case, for $R_{\rm H-H}$= 3.0 a.u., the system is strongly correlated and all Hartree-Fock orbitals used in simulations are non-negligible. This means that one cannot define a single small-dimensionality active space to capture all needed correlation effects for the $R_{\rm H-H}$= 3.0 a.u. case. Recently, the H$_n$ models have been  used for validation of  cutting-edge many-body numerical methodologies for treating correlated quantum systems \cite{PhysRevB.94.245129,motta2017towards,pfau2020ab,stair2020exploring,magoulas2022addressing}.

\begin{figure}
    \centering
    \includegraphics[width=0.34\textwidth]{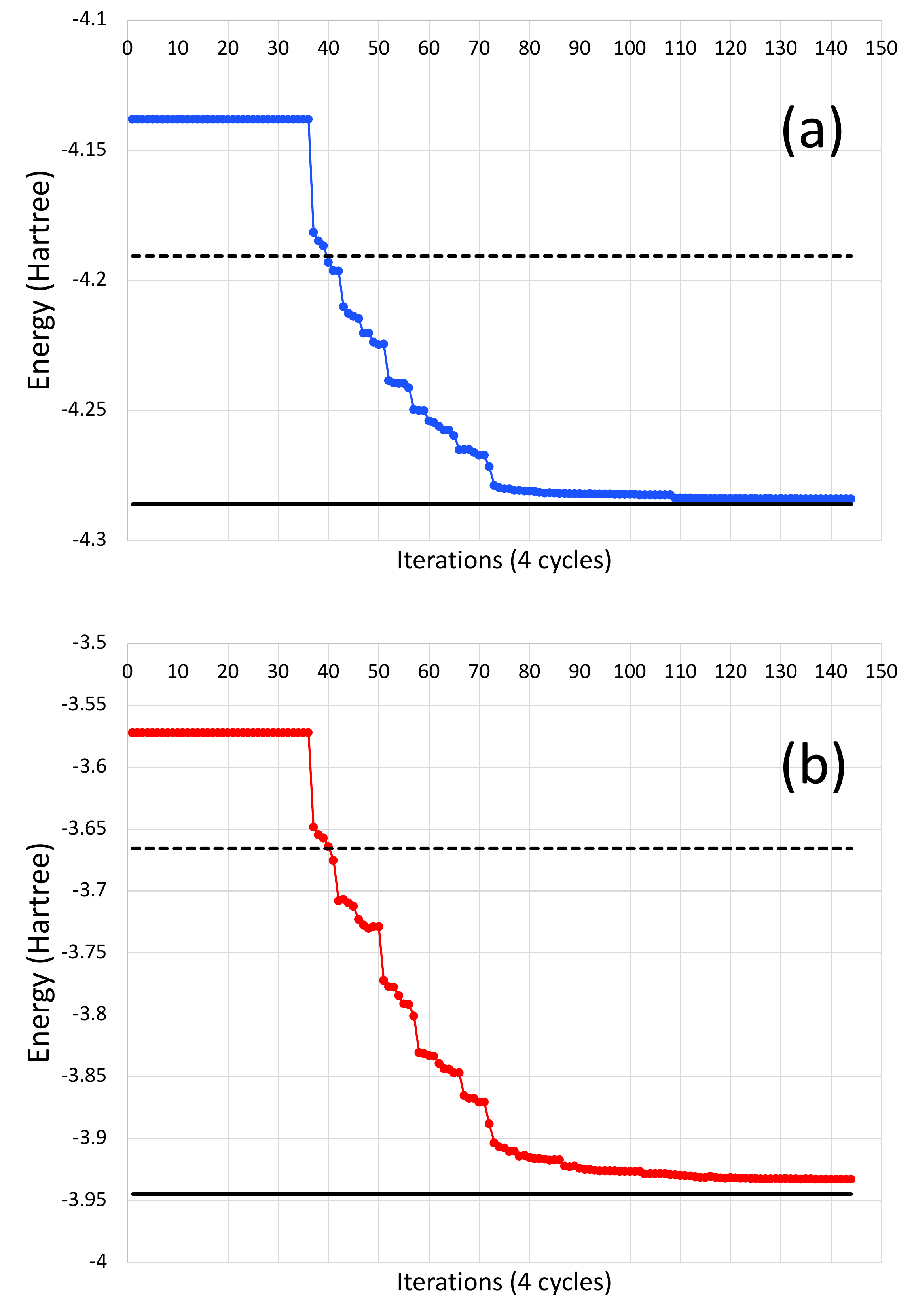}
    \caption{Energies evaluations (for all active spaces) in the QFlow for STO-3G H8 model: (a) $R_{\rm H-H}$ =2.0 a.u., blue circles, (b)  $R_{\rm H-H}$ =3.0 a.u., red circles. The dotted and solid horizontal black lines correspond to the active-space  and full-space exact diagonalizations, respectively. 
    We report energies of all active space problems for the first four cycles 
    To initiate the optimization process, we utilized a zero vector as the initial guess for all active spaces. The optimization process is based on the gradients (\ref{ducc6}), and the update of the parameter pool starts from the second cycle. Thus, all energies in the first cycle correspond to HF energies.
     }
    \label{fig:opt}
\end{figure}

\begin{figure}
    \centering
    \includegraphics[width=0.45\textwidth]{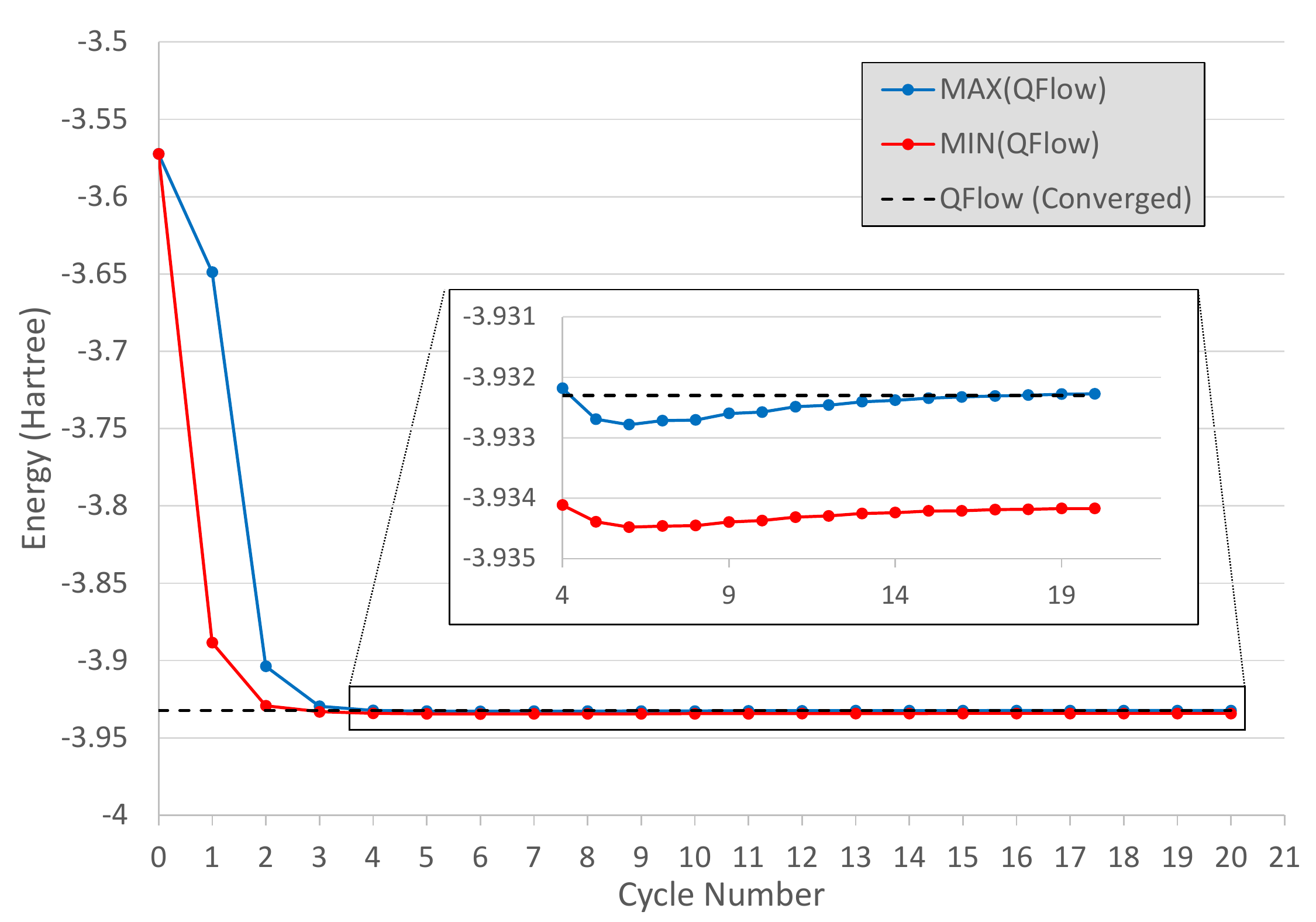}
    \caption{The minimum and maximum values of $E(\mathfrak{h}_i)$ at the beginning of each QFlow cycle of the STO-3G H8 $R_{\rm H-H}$=3.0 a.u. model.
}
    \label{fig:minmax}
\end{figure}

We summarized QFlow results in Table \ref{tab5} and in Figs. (\ref{fig:opt}) and (\ref{fig:minmax}).
For both systems, the QFlow included all active spaces defined by arbitrary two occupied active and   two virtual active orbitals and four active electrons (the QFlow(4e,4o) model). For H6 and H8 systems QFlow integrates 9 and 36 active spaces, respectively. In Table \ref{tab5}, the QFlow(4e,4o) results are compared against exact diagonalization (ED) in the full space,  in the primary active space (CAS-ED) consisting of the two highest energy occupied orbitals and two lowest energy unoccupied orbitals, and typical CC approximations including excitations from singles to quadruples (CCSD, CCSDT, and CCSDTQ) \cite{bartlett_rmp}.
It is evident that the QFlow algorithm significantly reduces errors of the CAS-ED method - a prevailing model for performing quantum simulations on NISQ-type devices.  In the extreme case, the error of CAS-ED amounting to  279 mHartree for the H8 3.0 a.u. system is reduced by QFlow to 12.4 mHartree. Additionally, it should be noticed that for the H6 and H8 $R_{\rm H-H}$=3.0 a.u. models,  the CCSD and CCSDT formulations experience  variational collapse placing the ground-state energies significantly below the ED ones.  
For weakly correlated H6 and H8 models ($R_{\rm H-H}$=2.0 a.u), the QFlow results are within chemical accuracy error bars (less than 1.59 mHartree). In Fig.~\ref{fig:opt}, we show energies ($E(\mathfrak{h}_i)$) calculated in the first four QFlow cycles for two geometries of H8. 
In both cases, we can observe that energies obtained  in the first non-trivial cycle (second cycle) are considerably better than the CAS-ED energy for the primary active spaces (targeted in typical VQE simulations). 
% revision
%{\color{blue}
%Further improvement of the QFlow performance, excluding the expansion of active spaces within the flow, can be attained by utilizing higher-order Trotter formulas to define
%$H^{\rm eff}(\mathfrak{h}_i,N)$ in Eq.(\ref{var3}).
%For example, for H8 $R_{\rm H-H}$=3.0 a.u. QFlow yields energy of -3.9350 Hartree, which effectively suppresses the QFlow error in comparison to the ED energy to within milliHartree.
%}
%end of revision
% revision: get rid of the following text
%{\color{red}
%The QFlow procedures with four active  electrons/four active orbital spaces limit the total pool of parameters, accounting for 
%differences between the converged QFlow energies and those obtained from the full ED. For example, higher-order excitations involving three or more different occupied or unoccupied orbitals cannot be accounted for because the active space does not allow it, having only two occupied and two unoccupied orbitals. The QFLow accuracies can be improved upon in two ways. The first is with larger active spaces, which will include a larger number of parameters/excitations. The other option is to use the spin-orbital definition of the active spaces (see Ref.~\cite{kowalski2023sub}) that enable the inclusion of broader classes of three- and four-body correlation effects without increasing computational cost per active space.
%end of revision

In Fig.~\ref{fig:minmax}, we discuss the discrepancies between the minimum and maximum values of $E(\mathfrak{h}_i)$ for each cycle for H8 3.0 a.u. model. Despite the fact that in cycles two and three these discrepancies are substantial, in the following iterations, these discrepancies significantly decrease. For 20-th cycle, the discrepancy is less than 2.0 mHartree, which 
indicates that despite approximations associated with  the non-commutative characters of cluster operators in QFlow,  the energy invariance of the  SR-CC flow (\ref{flow1}) at the solution is approximately satisfied. 
% revision
%{\color{blue}
%This discrepancy is further reduced to the 0.2 mHartree by order-3 Trotter expansion for the effective Hamiltonians (\ref{var6}).
%}
% end of revision
In the QFlow simulations for the H8 system, we optimized 684 parameters using coupled computational blocks corresponding to active space eigenvalue problems that optimize at most 35 parameters.
% revision
%{\color{blue}
Concurrently, the number of QFlow optimized amplitudes is considerably lower than that of CCSDTQ ones. For instance, QFlow optimizes only 36 quadruply excited amplitudes, while the CCSDTQ approach utilizes 1810 of them (with no spatial symmetry invoked in both techniques).
%}
% end of revision

%offset the errors of the CAS-ED procedure (a prevailing model for the NISQ-type quantum simulations)

% Advantage of VQE - it provides wave function parameters
% Future directions: hardware efficient representation of variational trial function & back cluster analysis. 
% --OK-- in \sigma_ext - also UCC parametrization.
\renewcommand{\tabcolsep}{0.2cm}
\begin{table}
    \centering
       \caption{Converged QFlow energies 
       (in Hartree) 
       for H6 and H8 benchmark systems at $R_{\rm H-H}$=2.0 a.u. and $R_{\rm H-H}$=3.0 a.u. corresponding to weakly and strongly correlated regimes, respectively.}
    \begin{tabular}{l c c c c }
    \hline \hline \\[-0.2cm]
%    \toprule
  Method & H6  & H6  & H8 &
  H8   \\
  &  ($2.0$ a.u) & ($3.0$ a.u) & ($2.0$ a.u) &
   ($3.0$ a.u) \\
  \hline \\[-0.2cm]
   HF           & -3.1059 & -2.6754 & -4.1382 & -3.5723    \\[0.1cm]
   CAS-ED        & -3.1669    &  -2.8021    &   -4.1906   &  -3.6656    \\[0.1cm]
   CCSD         & -3.2173  & -2.9673 & -4.2848 & -3.9727  \\[0.1cm]
   CCSDT        & -3.2180 & -2.9692 & -4.2867 & -3.9784 \\[0.1cm] 
   CCSDTQ       & -3.2177 & -2.9574 & -4.2860 & -3.9439 \\[0.1cm]
   QFlow(4e,4o)\footnote{QFlow energies are reported from the primary active space consisting of the two highest energy occupied orbitals and two lowest energy unoccupied orbitals. } & -3.2173    & -2.9521  & -4.2847 & -3.9322 \\[0.1cm]
   ED           & -3.2177 & -2.9576 & -4.2860 &  -3.9447   \\[0.1cm]  
         \hline \hline
    \end{tabular}
    \label{tab5}
\end{table}
% ---
%
%
%
%
%\begin{figure*}[!ht]
%\end{figure*}
%
%
%\begin{figure*}[!ht]
%\end{figure*}
%
%
\renewcommand{\tabcolsep}{0.2cm}
\begin{table}
    \centering
       \caption{The DUCC-QFlow energies (in Hartree) for the H6 model ($R_{\rm H-H}$=3 a.u.) in the  cc-pVDZ basis set.}
    \begin{tabular}{l c   }
    \hline \hline \\[-0.2cm]
%    \toprule
  Method &  Energy \\
  \hline \\[-0.2cm]
CCSD   &   -3.1571   \\[0.1cm]
CCSDT  &   -3.1619   \\[0.1cm]
CCSDTQ &   -3.1591   \\[0.1cm]
DUCC(6act)-QFlow(4e,4o) & -3.1570  \\[0.1cm]
DUCC(7act)-QFlow(4e,4o) & -3.1589  \\[0.1cm]
ED  & -3.1591  \\[0.1cm]
         \hline \hline
    \end{tabular}
    \label{tab6}
\end{table}
%
% revision
%{\color{blue}
While the STO-3G basis set is useful for validating the QFlow algorithm, in practical applications, larger basis sets that properly capture short-range dynamical correlation effects are required.  To address this challenge, we have implemented  a two-step strategy, referred to as the DUCC-QFlow approach described in Ref.\cite{kowalski2021dimensionality}. This approach utilizes (i) classical computers and a simplified downfolding technique to evaluate an approximate form of the effective Hamiltonian ($A$) for active spaces that are too large for current quantum hardware, and (ii) quantum computers to solve the problem described by Eq.(\ref{var1}) with $H$ replaced by $A$ using the QFlow algorithm. We illustrate  the feasibility of the DUCC-QFlow algorithm in handling larger basis sets on the challenging example of H6 for $R_{\rm H-H}$=3 a.u. for cc-pVDZ basis set
\cite{dunning1989gaussian}
and active orbitals  defined by six (DUCC(6act)-QFlow(4e,4o)) and seven (DUCC(7act)-QFlow(4e,4o)) lowest Hartree-Fock orbitals. As a downfolding procedure for the first step, we adopted the A(7) approximation for the  downfolding technique of Ref.\cite{doublec2022}. 
The DUCC-QFlow results shown in Table II indicate that while DUCC(6act)-QFlow(4e,4o) provides accuracies of the CCSD energies, the DUCC(7act)-QFlow(4e,4o) furnishes energies in a good agreement with the CCSDTQ or ED results. 
%}
% end of revision
%
%
%
%
\paragraph{Summary.---} We provided numerical evidence that the QFlow algorithm can efficiently sample large sub-spaces of the Hilbert space through coupled variational problems in reduced
dimensionality active spaces. Using very modest active space sizes, we illustrated the utility of the QFlow procedure with the STO-3G H6 and H8 hydrogen chains in weakly and strongly correlated regimes with errors within chemical accuracy for weakly correlated systems and relatively small errors for the strongly correlated systems.  For the strongly correlated H8 model, we recover nearly 97\% of the correlation using active spaces containing 
small number
%at most 5\% of the number
of optimized parameters compared to the exact diagonalization.
% revision
%{\color{blue} 
Additionally, the application of the two-step DUCC-QFlow protocol successfully accounted for correlation effects in the highly correlated version of the H8 molecule utilizing a larger cc-pVDZ basis set. Our expectations are that the DUCC-QFlow algorithm will facilitate the seamless integration of classical and quantum computational resources.
%}
% end of revision

The examples in this paper are very conservative estimates of the dimensionality reduction that can be achieved with the QFlow algorithm. As quantum technology evolves and we transition from the noisy intermediate-scale quantum devices era to fully-fledged error-corrected quantum computing, the ability to adapt to new methodological advances and efficiently utilize hybrid computational resources is ever-important. 
%The authors expect that the QFlow algorithm demonstrated in this letter will play a crucial role in expanding the envelope of many-body applications as quantum computing continues to evolve.
An intriguing aspect of QFlow, to be explored in forthcoming studies, is the possibility to leverage distributed quantum computing resources, as well as the ease with which QFlow can capture the local correlation effects using reduced number of qubits.
The authors expect that the QFlow algorithm demonstrated in this letter will play a crucial role in expanding the envelope of many-body applications as quantum computing continues to evolve.

% The progress in enabling quantum computing technologies is contingent not only on the advances in the design of quantum materials but also on the ability to adapt to new methodological advances in the theory of correlated many-body systems. Among the desired features of these methodologies, one should list the possibility of utilizing hybrid computational resources integrating the most appealing features of quantum and classical computing and the ability to construct flexible workflows that can naturally adapt to ever-evolving quantum computing technologies and provide much-needed tools for transitioning from noisy intermediate-scale quantum (NISQ) devices era to fully fledged error-corrected quantum computing. 

This material is based upon work supported by the ``Embedding QC into Many-body Frameworks for Strongly Correlated Molecular and Materials Systems''  project, which is funded by the U.S. Department of Energy, Office of Science, Office of Basic Energy Sciences, the Division of Chemical Sciences, Geosciences, and Biosciences (under FWP 72689) and by Quantum Science Center (QSC), a National Quantum Information Science Research Center of the U.S. Department of Energy (under FWP  )76213. This work used resources from the Pacific Northwest National Laboratory (PNNL).
PNNL is operated by Battelle for the U.S. Department of Energy under Contract DE-AC05-76RL01830.

%\clearpage

\bibliography{ref}
\end{document}